\documentstyle[11pt,newpasp,twoside]{article}
\index{instructions}
\index{guidelines}
\def\be{\begin{equation}}
\def\ee{\end{equation}}

\def\edcomment#1{\iffalse\marginpar{\raggedright\sl#1\/}\else\relax\fi}
\reversemarginpar

\begin{document}
\title{On the Bottom Magnetic Fields of the Millisecond Pulsars }
 \author{Chengmin Zhang}
\affil{  School  of  Physics,
       The University of Sydney,
       NSW 2006, Australia}

\begin{abstract}

The magnetic field strengths of most millisecond pulsars(MSP) are about
$10^{8-9}$ Gauss.
 The accretion induced 
 magnetic field evolution scenario here concludes that the field decay is 
invesely related to the accreted mass and the minimum field or bottom field 
stops at  about $10^{8}$  Gauss if accreted with the  Eddington accretion rate, which is 
proportionally related with the 
accretion rate as $\dot{M}^{1/2}$.
The  possibility  of the  
low field $\sim 10^{7}$ Gauss MSPs has been proposed 
 for the future radio observation.
\end{abstract}

As known, 
the low magnetic field (MF) and fast spinning of MSPs are ascribed to the 
mass accretion in the binary system.  The accretion induced neutron star 
magnetic field (NSMF) decay models 
have been proposed and 
paid much attention by a couple of authors (Romani 1990; 
Urpin \& Geppert 1995;  
Cheng \& Zhang 1998; Brown \& Bildsten 1998; 
Payne \& Melatos 2003), which is based on the observational evidence 
that NSMF decays in the binary accretion phase (Taam \& van den Heuvel 1986; 
Shibazaki et al. 1989). Moreover, Van den Heuvel and
Bitzaraki (1995)  discovered that 
 the increased amount of mass accreted leads to decay of the 
 NSMF, and the 'bottom' field strength of about 10$^8$ Gauss is also   
implied.

 Based on the accretion induced field decay in the slab geometry approximation 
proposed by Cheng \& Zhang (1998), we extend it to the 
the  spherical geometry structure of the accreted neutron star
illustrated in Fig.1 of paper by  Cheng \& Zhang (1998). 
We assume that the magnetic field lines are 
frozened in
the entire NS crust with homogenous average mass density.
 Under the condition of  incompressible fluid approximation as well as the
constant crustal volume assumption, the piled accreted materials will arise
the expansion of the volume of the polar zone by 
$\delta V_p =\frac{\dot{M} \delta t}{\rho} - A_p\delta H $, 
where $A_p$ is the surface area of the polar zone, $\rho$ is the average density
of the crust, $\dot{M}$ is the accretion rate, $\delta t$ a particular accretion
duration, H the thickness of the crust and $\delta H$/H the fraction of the
thickness dissolving into the core which is defined by 
$\frac{\delta H}{H} = \frac{\dot{M} \delta 
t}{M_{cr}}$, where $M_{cr}=4\pi R^2 \rho H$ is the crust mass.
 The expansion of the polar zone
will dilute the magnetic flux density if the conservation of the magnetic flux
is preserved, so, $\delta(BA_p)=0$, 
 and the area $A_p$ of the accretion polar patch can be accurately expressed as
follows(Shapiro \& Teukolsky 1983),
$A_p = 2\pi R^2 (1-cos\theta_c)\,\,  , \,\, sin^2\theta_c =\frac{R}{R_A}$, 
where $\theta _c$ is the 
angle between field line of the star surface and polar
axis, $R_A$ is the Alfven radius, 
$R_A =3.2\times 10^8 \,{(cm)}\, 
\dot{M}^{-2/7}_{17} \mu^{4/7}_{30} (\frac{M}{M_\odot})^{-1/7}$, 
 where M is the neutron star mass, 
$\dot{M}_{17}$ is the accretion rate in
units of $10^{17} g/s$, and $\mu _{30}$ is the magnetic moment in unit
of $10^{30} G cm^3$. Therefore,  using the specified relation
$\delta V_p =H\delta A_p$ and the equation of conservation of magnetic flux,
we get the following relation, $\delta V_p = -V_p \frac{\delta B}{B}$. 
 Connecting  the equations above,  we obtain 
 MF  evolutionary equation as,
${A_p\delta B}/{[(2\pi R^2 - A_p)B]}={\dot{M}\delta t}/{M_{cr}}$. 
Considering    the initial condition B(t=0)=$B_0$, we have,\\
$B/B_f = \{1 - [C{}\exp(-\frac{2\Delta M}{7M_{cr}})-1]^2\}^{-7/4}\,,
C = 1+\sqrt{1-(\frac{B_f}{B_0})^{4/7}},$\\ 
$\Delta M = \dot{M} t$ and $B_f$ is the magnetic field defined by the Alfven
radius matching the star radius, i.e., $R_A (B_f)=R$, which is also the minimum 
field strength, or named as the bottom field (van den Heuvel \& Bitzaraki 1995), 
\\$B_f = 1.32 \times 10^8 (\frac{\dot{M}}{\dot{M}_{Ed}})^{1/2}
(\frac{M}{M_\odot})^{1/4}R^{-5/4}_6 \,\,\,(G),$\\
where $\dot{M}_{Ed}$ is the Eddington accretion rate 
and $R_6$ is NS radius
  in units of $10^6$ cm. 
If $\Delta M = 10^{-5}\sim 10^{-4} M_\odot$,
MF  evolutionary equation can be simplified as the  following
approximated form,
$B = {B_0}/(1 + \frac{\Delta M}{m_B})$, 
where $m_B = \frac{1}{2}(B_f/B_0)^{4/7}M_{cr}\simeq
5 \times 10^{-4} (M_\odot) (M_{cr}/0.2 M_{\odot}) $.
This  is just the same form as the empirical formula
of accretion induced field decay proposed by Shibazaki et al. (1989).

The main conclusions  are summaried  
in the following. NSMF decays in the binary accretion phase, which 
is inversely correlated with the mass 
acreted from the companion. 
The bottom  field of MSP is determined by the condition of 
that the magnetosphere radius equals the star radius, which is proportionally 
related to the accretion rate as $\dot{M}^{1/2}$ (White \& Zhang 1997). 
The bottom field of Z-source 
(Eddington  accretion rate) is about 
$\sim 10^{8}$ Gauss, and the  bottom field of Atoll-source (1\% Eddington 
 accretion rate) will correspond to  $\sim 10^{7}$ 
Gauss.    The so low field $\sim 10^{7}$ Gauss MSPs have not yet been 
discovered from the  radio observation. 
The final state of NS is constrained by the system parameters, such as 
 star radius, star mass and mass accretion rate, which is nothing to do 
with the initial field and initial period, so this is why the almost homogeneous 
field 
distributions of MSPs from  $\sim 10^{8}$ (G) to  $\sim 10^{9}$ (G) do not follow 
the 
field distributions of the normal PSRs  from  $\sim 10^{11}$ (G) to  $\sim 10^{15}$ (G).



\end{document}